\documentstyle[epsfig]{elsart}

\begin{document}

\begin{frontmatter}

\title{Anatomy of three-body decay \\
III. Energy distributions}

\author{E. Garrido} 
\address{ Instituto de Estructura de la Materia, CSIC, 
Serrano 123, E-28006 Madrid, Spain }

\author{D.V. Fedorov, A.S.~Jensen \and H.O.U. Fynbo}
\address{ Department of Physics and Astronomy,
        University of Aarhus, DK-8000 Aarhus C, Denmark }

\date{\today}

\maketitle

\begin{abstract}
We address the problem of calculating momentum distributions of
particles emerging from the three-body decay of a many-body resonance.
We show that these distributions are determined by the asymptotics of
the coordinate-space complex-energy wave-function of the resonance.
We use the hyperspherical adiabatic expansion method where all lengths
are proportional to the hyperradius.  The structures of the resonances
are related to different decay mechanisms.  For direct decay all
inter-particle distances increase proportional to the hyperradius at
intermediate and large distances.  Sequential three-body decay
proceeds via spatially confined quasi-stationary two-body
configurations. Then two particles remain close while the third moves
away. The wave function may contain mixtures which produce coherence
effects at small distances, but the energy distributions can still be
added incoherently. Two-neutron halos are discussed in details and
illustrated by the $2^+$ resonance in $^{6}$He. The dynamic evolution
of the decay process is discussed.
\end{abstract}

\end{frontmatter}

\par\leavevmode\hbox {\it PACS:\ } 21.45.+v, 31.15.Ja, 25.70.Ef

\maketitle

\section{Introduction}

Nuclear resonances and excited states can be very complicated
many-body structures with a number of different decay modes.  The
simplest decay, perhaps beside $\gamma$-emission, is breakup into two
particles as exemplified by nucleon- and $\alpha$-emission, and binary
fission \cite{sie87}. The deceivingly simple breakup into three
particles is much less studied and far from understood. This is in
contrast to bound three-body cluster structures where a variety of
techniques are available and able to predict the properties, even of
the exotic quantum halo states \cite{jen04}.  The three-body continuum
properties are less established although rather well studied over many
years \cite{glo96}.

Experimental information is obtained by measuring the properties of
the particles in the final state. The experimental techniques are now
advanced to a level where accurate and kinematically complete
measurements are available on a number of different systems
\cite{dan87,boc89,kry95,bai96,gom01,gio02,pfu02,chr02,fyn03,bla03,zer04,bla05},
and many more are expected to follow. Reliable theoretical
descriptions are needed to interpret existing data, to predict
unknown decay results and to help in the design of new interesting
experiments.  Both the structure of the initial state and the decay
mechanism are essential and both must therefore be properly described
simultaneously.  Clearly for a genuine many-body state only the
intermediate and large-distance structure is decisive.  The small
distance behavior is artificial and only serving to provide the proper
continuous boundary conditions.

For two-body decay, like $\alpha$-emission, the relative potential
determines all properties.  In the example of $\alpha$-emission, the
two-body potential can be divided into short-, intermediate- and
long-distance.  The short-distance part is artificially adjusted to
allow the correct resonance energy and the barrier region then
determines the width.  At large distances where the potential has
vanished the energy of the $\alpha$-particle is determined by energy
conservation. In two-body decay the energy distribution then only
reflects the width of the initial resonance.  These properties are
very different for decays with more than two particles in the final
state.

For three-body bound states and resonances ``large distance'' is less
well defined . However, the corresponding structure can efficiently be
computed by use of the hyperspherical adiabatic expansion method
\cite{nie01,fed02}.   The wave functions are in this technique expanded
on basis states related to adiabatic potentials calculated as
functions of the hyperradius $\rho$.  Then $\rho$ provides a measure
of distances for the three-body problem.  The wave function is usually
dominated by the component related to the lowest adiabatic potential.
The small-distance part ($\rho$ small) of both wave function and
potential are only directly meaningful if the particles appearing in
the final state form a genuine three-body system. Otherwise this part
of the effective potential is constructed to produce the correct
resonance energy and provide an appropriate boundary condition for the
wave function.  At intermediate distances the potential has a barrier
which is decisive for the width of the resonance.  At large distances
the resonance wave function is characterized by outgoing waves which
contain information about distributions of relative energies and
possibly other quantities like spin distributions.

For three-body decay the two most obvious decay mechanisms, direct and
sequential, were recently studied in a schematic model in
\cite{gar05a}.  Another schematic model is also formulated in the limit 
where only the Coulomb interaction is important at intermediate and
large distances \cite{kar04}.  The detailed resonance
structure at small and intermediate distances were investigated in
realistic models in \cite{gar05b}.  The large-distance properties are
much more difficult to calculate accurately, because either the
correct continuum three-body Coulomb wave functions are unknown, or the
short-range potentials may produce an almost long-range (inverse
square) effective potential at large distance.  In the latter case the
origin is precisely as for the Efimov effect \cite{nie01,gar05c}. The
corresponding potential is most likely the lowest at large distance
but not necessarily also at small distances.  Thus, the resonance
wave function may change structure from small to large distance. The
relative energy distribution arises as the result appearing at large
distance.  The numerical computations are then sometimes rather
tricky.

The purpose of the present paper is to establish a general method to
compute relative energy distributions after decay into three
particles.  The short- and intermediate-distance resonance structure
from \cite{gar05a,gar05b} is a good starting point but we need in
addition to calculate the asymptotic behavior in momentum space.  The
asymptotics vary for the different decay mechanisms, and the related
numerical treatment is difficult when all the possibilities
simultaneously have to be accounted for.

We assume that formation and decay of the resonances are independent
processes.  The resonance could be formed by beta-decay from a
neighbouring nucleus, or a window with the relevant energies can be
selected in an experimental setup where contributions from other
processes also are eliminated. In section 2 we develop the theoretical
formalism for resonance decay. This was previously sketched by use of
the saddle point approximation \cite{fed04}, while we here shall
instead use the Zeldovich regularization of the divergent Fourier
integrals \cite{zel60}. In section 3 we discuss some of the important
features arising from calculation of resonance wave functions by use of
the hyperspherical adiabatic expansion method combined with the
complex scaling method.  In section 4 we illustrate in details with
realistic computations for the $2^+$-resonance in $^{6}$He.  Finally,
section 5 contains a brief summary and the conclusions.

\section{Theoretical formulation}

We assume that the system of particles has been generated in a meta-stable
quantum state (a resonance) that is a generalized eigen-state of the
Hamiltonian with complex energy. This is a decaying state -- it describes a
constant flux of particles towards infinity. Suppose we have a system of
detectors at large distances which measure the momenta of the particles
emerging from this decaying state. Clearly these detectors will measure the
probability distribution of particle momenta in the meta-stable state,
that is the absolute square of the momentum space wave-function.

\subsection{Two-body resonances}

The theoretical derivation is most easily understood if we first
explain the idea for simple resonance decay into a two-body system.
We need resonance inventions in coordinate and momentum-space and
transformations between these non-square integrable functions.

\subsubsection{Resonance wave functions}

The momentum space wave-function of a resonant state with the complex energy
$E_{r}=\frac{\hbar ^{2}}{2m}k_{0}^{2}=E_{0}-i\frac{\Gamma }{2}$ has the form
\cite{new?}
\begin{equation}
\psi _{k_{0}}({\mathbf{k}}) = \frac{g(k,\Omega _{k})}{k^{2}-k_{0}^{2}} \; ,
\label{psi_k}
\end{equation}
where $\mathbf{k}$ is the relative momentum and $\Omega _{k}$
indicates the two angles defining the direction of the vector
$\mathbf{k}$. We assume that the wave function $\psi
_{k_{0}}(\mathbf{k})$ only has the pole at $k = k_{0}$ and $g(k,\Omega
_{k})$ is then a continuous function of the momentum $\mathbf{k}$ with
no poles.

The distribution $P(\mathbf{k)}$ of the relative momentum $\mathbf{k}$ of
the two particles in the resonant state $\psi _{k_{0}}$ is given by the
absolute square of the momentum-space wave function, i.e.
\begin{equation}
P({\mathbf{k}})=|\psi _{k_{0}}({\mathbf{k}})|^{2} \propto\frac{|g(k,\Omega
_{k})|^{2}}{(E-E_{0})^{2}+\frac{\Gamma ^{2}}{4}} \; ,
\end{equation}
where the real observable energy $E$ is $E=\hbar ^{2}k^{2}/(2m)$. The
approximation that the system is generated in a pure resonant state
$\psi _{k_{0}}$ is most likely only valid in the neighborhood of the
resonant energy, i.e.  $E \simeq E_{0}$. Furthermore, the function
$g(k,\Omega _{k})$ is smooth and varies by definition much less than
the denominator in eq.(\ref{psi_k}).  In any case we shall only
consider energies where $|E-E_0|$ is less than a few times $\Gamma$.
We can then confidently substitute the momentum $k$ with the
resonant momentum $k_{0}$ in $g(k,\Omega _{k})$ and thus arrive at the
expression of the famous Breit-Wigner type
\begin{equation}
P({\mathbf{k}})\propto \frac{|g(k_{0},\Omega _{k})|^{2}}{(E-E_{0})^{2}+\frac{
\Gamma ^{2}}{4}} \; ,  \label{p_k}
\end{equation}
where the energy dependence is given by the factor $\left[
(E-E_{0})^{2}+\frac{\Gamma ^{2}}{4}\right] ^{-1}$ while the angular
dependence is given by the (absolute square of the) function
$g(k_0,\Omega _{k})$. Thus, the momentum-space wave-function
eq.(\ref{psi_k}) of the resonance allows direct calculation of the
momentum distributions of the decay fragments through eq.(\ref{p_k}).
Clearly improvements are possible by use of different approximations
of $g(k,\Omega _{k})$.

Instead of momentum-space, the wave-function of the resonance may be
available in coordinate-space where the large-distance asymptotic form
is given by
\begin{equation} \label{e60}
\psi _{k_{0}}({\mathbf{r}})\stackrel{r\rightarrow \infty }{\longrightarrow }
\frac{e^{+ik_{0}r}}{r}f(\Omega _{r}) \; ,
\end{equation}
where $\Omega _{r}$ denotes the two angles defining the direction of
the relative coordinate ${\mathbf{r}}$.  The structure is an outgoing
spherical wave potentially modified by an angular dependence contained
in $f(\Omega _{r})$. Generally the resonance wave function can be
written as a partial-wave expansion in the spherical harmonics
$Y_{lm}$, i.e.
\begin{equation}
\psi _{k_{0}}({\mathbf{r}})=\sum_{lm}C_{lm}\chi _{l}(r)Y_{lm}(\Omega _{r}) \; ,
\end{equation}
where $C_{lm}$ are constants depending on angular momentum and
projection quantum numbers $l$ and $m$. The radial functions $\chi
_{l}(r)$ are those solutions of the radial Schr\"{o}dinger equation
that asymptotically approach the outgoing spherical wave in
eq.(\ref{e60}), i.e. 
\begin{equation}   \label{chi-}
\chi _{l}(r)\stackrel{r\rightarrow \infty }{\longrightarrow }\frac{
e^{+ik_{0}r}}{r} \;\; ,\;\,
f(\Omega _{r})=\sum_{lm}C_{lm}Y_{lm}(\Omega _{r}) \; .
\end{equation}

This defines the asymptotic behavior of the decaying resonance
wave function which in turn determines the energy distribution in the
observable final state.

\subsubsection{Transformation from coordinate- to momentum-space}

The coordinate-, $\psi _{k_{0}}(\mathbf{r})$, and momentum-space, $
\psi _{k_{0}}(\mathbf{k})$, wave-functions are connected via a Fourier
transform
\begin{equation} \label{e70}
\psi _{k_{0}}({\mathbf{k}})=\int e^{-i{\mathbf{kr}}}\psi _{k_{0}}
({\mathbf{r}})d^{3}r.
\end{equation}
Expansion of the plane-wave in terms of spherical harmonics
\begin{equation}
e^{i{\mathbf{kr}}}=\sum_{lm}4\pi i^{l}j_{l}(kr)Y_{lm}(\Omega _{r})
Y_{lm}^{*}(\Omega _{k})
\end{equation}
reduces the Fourier integral in eq.(\ref{e70}) to a one-dimensional
radial integral
\begin{equation} \label{e75}
\psi _{k_{0}}({\mathbf{k}})=4\pi \sum_{lm}C_{lm}(-i)^{l}Y_{lm}(\Omega _{k})
\int_{0}^{\infty }r^{2}dr\chi _{l}(r)j_{l}(kr) \; .
\end{equation}
Because of the asymptotics in eq.(\ref{chi-}) the radial integral is
seen to diverge.  The large-distance behavior is responsible for the
divergence.  The physics content, expressed by a finite value, then
has to be extracted by a suitable regularization.  We use the
prescription proposed by Zeldovich \cite{zel60}, i.e. multiplication
of the integrand by a gaussian where the range after integration is
increased to infinity.  For an exponential this gives
\begin{equation} \label{e77}
\int_{0}^{\infty }e^{iqr}dr\rightarrow \lim_{\alpha \rightarrow
0}\int_{0}^{\infty }e^{iqr-\alpha ^{2}r^{2}}dr=\lim_{\alpha \rightarrow
0}e^{-\frac{q^{2}}{4\alpha ^{2}}}\frac{\sqrt{\pi }}{2\alpha }
{\mathrm{erfc}}(- \frac{iq}{2\alpha })=\frac{i}{q} \; ,
\end{equation}
where $q$ can be complex and $erfc$ is the error function of complex
argument.  In the present context the radial integral in
eq.(\ref{e75}) is first rewritten by adding and subtracting the
asymptotic expression of the diverging integrand. The difference
between the true and the asymptotic expression then remains finite
even without multiplication by the gaussian function.  Only the
asymptotic expression then diverges when the gaussian smoothly
converges to an overall factor of one.

The physics content is extracted by dividing with a similarly
diverging normalization integral of the square of the wave function
$\chi _{l}(r)$.  Also this integral, now in the denominator, is
rewritten by addition and subtraction of its asymptotic
expression. Again only the asymptotic expression diverges. The
Zeldovich prescription now leaves the ratio of these two diverging
integrals of the asymptotic expressions. However, this ratio does not
diverge but converge towards the physically meaningful result.  Apart
from a normalization we therefore have to regularize only the
asymptotic expression obtained by use of eq.(\ref{chi-}) and the
asymptotic approximation of $j_{l}(kr)$, i.e.
\begin{equation} \label{e85}
\psi _{k_{0}}({\mathbf{k}})=4\pi \sum_{lm}C_{lm}(-i)^{l}Y_{lm}(\Omega_{k})
 \int_{0}^{\infty }r^{2}dr\frac{e^{+ik_{0}r}}{r}\frac{\sin (kr-\frac{l\pi }{2
})}{kr} \; .
\end{equation}
The radial integral is then by use of eq.(\ref{e77}) evaluated to be
\begin{eqnarray}
\int_{0}^{\infty }e^{+ik_{0}r}\sin (kr-\frac{l\pi }{2})dr = \frac{
i^{l}}{2}\left[ \frac{1}{k-k_{0}}-\frac{(-1)^{l}}{k+k_{0}}\right]
\nonumber\\
=\frac{i^{l}}{2}\frac{k+k_{0}-(-1)^{l}(k-k_{0})}{k^{2}-k_{0}^{2}} =
 i^{l} \frac{k_{0}}{k^{2}-k_{0}^{2}} \;\; \;\; {\rm or } \;\;\;\;
 i^{l} \frac{k}{k^{2}-k_{0}^{2}} \; 
\end{eqnarray}
for even or odd $l$, respectively.  The summation in eq.(\ref{e85}) is
proportional to the angular amplitude $f$ from eq.(\ref{chi-}), but
now as a function of the momentum $\mathbf{k}$.  In any case we
assumed earlier that $k\approx k_{0}$ in the smooth functions.  We
therefore arrive at the final expression for the Fourier-transform of
the resonance wave-function, i.e.
\begin{equation}
\psi _{k_{0}}({\mathbf{k}})=\frac{4\pi }{k^{2}-k_{0}^{2}}
 \sum_{lm}C_{lm}Y_{lm}(\Omega _{k})=\frac{4\pi }{k^{2}-k_{0}^{2}}
 f(\Omega _{k}) \; .
\end{equation}
Thus the function $g$ from eq.(\ref{psi_k}) is then related to $f$
from eq.(\ref{e60}) by
\begin{equation}
g(k_{0},\Omega _{k})=4\pi f(\Omega _{k}) \; .
\end{equation}
The convenient fact that only the asymptotic limit of the resonance
wave function enters after the regularization procedure is perhaps more
surprising in mathematics than in physics where the observable energy
distributions always are obtained from the properties at large
distances.

The observable distribution in momentum-space is determined by the
angular wave function in coordinate-space evaluated for angles
describing the direction of the momentum.  This peculiar fact can
intuitively be understood by the geometry of particles moving towards
the detectors at infinitely large distances.  Coordinates and momenta
then must point in the same direction.  A mathematical formulation is
available from ionization cross sections calculated for atomic physics
processes \cite{ovc04}.

\subsection{Three-body resonances}

The generalization to three particles first requires a convenient set
of coordinates. We choose the scaled Jacobi coordinates \cite{nie01}
\begin{eqnarray}
{\mathbf{x}} &=&\sqrt{\frac{m_{2}m_{3}}{m(m_{2}+m_{3})}}({\mathbf{r}}_{2}-
{\mathbf{r}}_{3}), \\
{\mathbf{y}} &=&\sqrt{\frac{m_{1}(m_{2}+m_{3})}{m(m_{1}+m_{2+}m_{3})}}\left(
{\mathbf{r}}_{1}-\frac{m_{2}{\mathbf{r}}_{2}+m_{3}{\mathbf{r}}_{3}}
 {m_{2}+m_{3}} \right) ,  \nonumber
\end{eqnarray}
where $m$ is a mass scale, and ${\mathbf{r}}_{i}$ and $m_{i}$ are the
coordinate and mass of the particle number $i$. The hyper-spherical
coordinates are then the hyper-radius $\rho $, the hyper-angle $\alpha
$, and the directional angles $
\Omega _{x}$ and $\Omega _{y}$ of the vectors $\mathbf{x}$ and $\mathbf{y}$
\textbf defined by
\begin{equation}
\rho =\sqrt{x^{2}+y^{2}} \;\; , \;\;\alpha =\arctan (x/y) \;\;,
\;\;\Omega _{\rho }=\{\alpha ,\Omega _{x},\Omega _{y}\} \; .
\end{equation}
The Jacobi coordinates depend on the sequence chosen for the
particles, and the three different pairs of $\mathbf{x}$ and
$\mathbf{y}$ could be labeled to distinguish.  We omit these labels
when the meaning is clear.

The corresponding momentum-space variables are
\begin{equation}
\kappa =\sqrt{p^{2}+q^{2}}\;\; , \;\;\alpha _{\kappa }=\arctan (q/p)
 \;\; , \;\; \Omega _{\kappa}=\{\alpha ,\Omega _{p},\Omega _{q}\} \; ,
\end{equation}
where $\mathbf{p}$ and $\mathbf{q}$ are the conjugate momenta related
to the coordinates $\mathbf{x}$ and $\mathbf{y}$.

\subsubsection{No bound two-body subsystems}

The generalization of the two-body spherical harmonics are the so-called
hyper-spherical harmonics \cite{nie01}
\begin{eqnarray}
{{\mathcal{Y}}}_{{{\mathcal{K}}}}(\Omega _{\rho
})=N_{n}^{(l_{x},l_{y})} \sin^{l_{x}} \alpha  \cos^{l_{y}} \alpha
 P_{n}^{(l_{x}+\frac{1}{2},\;l_{y}+\frac{1}{2})}(\cos 2\alpha)
\nonumber\\
\times 
Y_{l_{x}m_{x}}(\Omega _{x})Y_{l_{y}m_{y}}(\Omega _{y})
\end{eqnarray}
where ${{\mathcal{K}}\equiv \{Kl_{x}m_{x}l_{y}m_{y}\}}$,
$K=2n+l_{x}+l_{y}$, and $(l_{x},m_{x},l_{y},m_{y})$ are the angular
quantum numbers related to coordinates $\bf{x}$ and $\bf{y}$, and
$N_{n}^{(l_{x},l_{y})}$ is a normalization factor.  These functions
are the eigen-functions of the angular part $\Lambda ^{2}$ of the
three-body kinetic energy operator $T$
\begin{equation}
T=\frac{\hbar ^{2}}{2m}(\nabla _{x}^{2}+\nabla _{y}^{2})=\frac{\hbar ^{2}}{2m
}\left[ -\frac{\partial ^{2}}{\partial \rho ^{2}}-\frac{5}{\rho }\frac{
\partial }{\partial \rho }+\frac{\Lambda ^{2}}{\rho ^{2}}\right]
\end{equation}
with the eigenvalues $K(K+4)$, i.e.
\begin{equation}
\Lambda ^{2}{\mathcal{Y}}_{{\mathcal{K}}}(\Omega _{\rho })=K(K+4)
{\mathcal{Y}}_{{\mathcal{K}}}(\Omega _{\rho }) \; ,
\end{equation}
where $K$ is a non-negative integer.  Without Coulomb and without
bound two-body subsystems the three-body resonance wave-function $\Psi
_{\kappa _{0}}(\rho ,\Omega _{\rho })$ with the complex energy
$E_r=\hbar ^{2}\kappa _{0}^{2}/(2m) = E_0 -i \Gamma_0/2$ can be
expanded in terms of the hyper-spherical harmonics
\begin{equation}
\Psi _{\kappa _{0}}(\rho ,\Omega _{\rho })=\sum_{{\mathcal{K}}}C_{{\mathcal{K}}
}\chi _{{\mathcal{K}}}(\rho ){\mathcal{Y}}_{{\mathcal{K}}}(\Omega
_{\rho }),
\label{eq21}
\end{equation}
where the hyper-radial functions $\chi _{{\mathcal{K}}}(\rho )$ have
the usual resonance asymptotic behavior of an out-going
hyper-spherical wave
\begin{equation} \label{e115}
\chi _{{\mathcal{K}}}(\rho )\stackrel{\rho \rightarrow \infty }{
\longrightarrow }\frac{e^{+i\kappa _{0}\rho }}{\rho ^{5/2}}.
\label{eq22}
\end{equation}
The three-body wave-function asymptotically has the form of the
hyper-spherical wave with an angular amplitude $F(\Omega _{\rho })$
determined by the expansion coefficients $C_{{\mathcal{K}}}$, i.e.
\begin{equation}\label{asy}
\Psi _{\kappa _{0}}(\rho ,\Omega _{\rho })\stackrel{\rho \rightarrow \infty
}{\longrightarrow }\frac{e^{+i\kappa _{0}\rho }}{\rho ^{5/2}}\sum_{{\mathcal{K
}}}C_{{\mathcal{K}}}{\mathcal{Y}}_{{\mathcal{K}}}(\Omega _{\rho })\equiv \frac{
e^{+i\kappa _{0}\rho }}{\rho ^{5/2}}F(\Omega _{\rho }).
\end{equation}
The momentum-space wave-function is the Fourier transform
\begin{equation} \label{e90}
\Psi _{\kappa _{0}}(\kappa ,\Omega _{\kappa })=\int e^{-i{\mathbf{px}}-i
{\mathbf{qy}}}\Psi _{\kappa _{0}}(\rho ,\Omega _{\rho })\rho ^{5}d\rho d\Omega
_{\rho }.
\end{equation}
The three-body plane-wave can be expanded in hyper-spherical harmonics as
\begin{equation}
e^{i{\mathbf{px}}+i{\mathbf{qy}}}=\frac{(2\pi )^{3}}{(\kappa \rho )^{2}}\sum_{
{\mathcal{K}}}i^{K}J_{K+2}(\kappa \rho ){\mathcal{Y}}_{{\mathcal{K}}}(\Omega
_{\rho }){\mathcal{Y}}_{{\mathcal{K}}}^{*}(\Omega _{\kappa }).
\end{equation}
Due to orthogonality of the hyper-spherical harmonics the angular part
of the integral in eq.(\ref{e90}) is trivial and we are only left with
the hyper-radial integral
\begin{equation}
\Psi _{\kappa _{0}}(\kappa ,\Omega _{\kappa })=\frac{(2\pi )^{3}}{\kappa ^{2}
}\sum_{{\mathcal{K}}}(-i)^{K}C_{{\mathcal{K}}}
{\mathcal{Y}}_{{{\mathcal{K}}}}(\Omega
_{\kappa })\int \rho ^{3}d\rho \chi _{{\mathcal{K}}}(\rho )J_{K+2}(\kappa \rho
).
\end{equation}
Precisely as in the two-body case, in the vicinity of the resonance
the integrand can be made ready for regularization by substitution of
eq.(\ref{e115}) and the asymptotic form
\begin{equation}
J_{K+2}(\kappa \rho )\stackrel{\rho \rightarrow \infty }{\longrightarrow }-
\sqrt{\frac{2}{\pi \kappa \rho }}\sin (\kappa \rho -\frac{\pi K}{2}) \;.
\end{equation}
This results in a diverging integral similar to that of the two-body
case
\begin{equation}
\Psi _{\kappa _{0}}(\kappa ,\Omega _{\kappa })=-\frac{(2\pi )^{3}}{\kappa
^{5/2}}\sqrt{\frac{2}{\pi }}\sum_{{\mathcal{K}}}(-i)^{K}C_{{\mathcal{K}}}
{\mathcal{Y}}_{{\mathcal{K}}}(\Omega _{\kappa })\int d\rho e^{+i\kappa _{0}\rho
}\sin (\kappa \rho -\frac{\pi K}{2}).
\end{equation}
Using the Zeldovich regularization leads to
\begin{eqnarray}
\Psi _{\kappa _{0}}(\kappa ,\Omega _{\kappa })=-\frac{(2\pi )^{3}}{\kappa
_{0}^{5/2}}\sqrt{\frac{2}{\pi }}\sum_{{\mathcal{K}}}(-i)^{K}C_{{\mathcal{K}}}
{\mathcal{Y}}_{{\mathcal{K}}}(\Omega _{\kappa })\frac{i^{K}}{2}\frac{2\kappa _{0}
}{\kappa ^{2}-\kappa _{0}^{2}}
\nonumber\\
=-\frac{2^{7/2}\pi ^{5/2}}{\kappa _{0}^{3/2}}
\frac{1}{\kappa ^{2}-\kappa _{0}^{2}}F(\Omega _{\kappa }),
\end{eqnarray}
that is, in the vicinity of the resonance, the angular wave function in
momentum-space is proportional to that in coordinate-space but
evaluated for the momentum variables. The energy distribution is
determined by the Breit-Wigner factor where the width is obtained from
the three-body resonance.  The function $F(\Omega _{\kappa })$ now
contains information about the non-trivial energy distribution between
the three particles.  This is in contrast to the two-body decay where all
the energy is in the only existing relative degree of freedom.

\subsubsection{One bound two-body subsystem}

Sometimes a bound two-body subsystem can be emitted from a three-body
resonance. Such a final state configuration can not be described by
hyper-spherical harmonics. However if this is the only open channel,
the description of such a decay reduces to the two-body case. Indeed
the asymptotics of the resonance wave-function is then
\begin{equation}
\Psi _{\kappa _{0}}(\rho ,\Omega _{\rho })\stackrel{\rho \rightarrow \infty
}{\longrightarrow }\phi _{23}({\bf{x}})\frac{e^{iq_{0}y}}{y}f(\Omega _{y}),
\end{equation}
where $\phi _{23}(\bf{x})$ describes a bound system of particles 2 and
3 with binding energy $B_{23}$, $q_{0}=\sqrt{2m(E_r-B_{23})/\hbar
^{2}}$, $f(\Omega _{y})$ is the angular amplitude and $E_r$ is the 
complex three-body energy.  If both three-body and two-body
decays are possible the wave-function contains asymptotics of both
two- and three-body types,
\begin{equation}
\Psi _{\kappa _{0}}(\rho ,\Omega _{\rho })\stackrel{\rho \rightarrow \infty
}{\longrightarrow }A\frac{e^{+i\kappa _{0}\rho }}{\rho ^{5/2}}F(\Omega
_{\rho })+B\phi_{23}({\bf{x}})\frac{e^{iq_0y}}{y}f(\Omega _{y}),
\end{equation}
where $A$ and $B$ are the asymptotic coefficients determining the
relative weights of the two decay channels.  Both $F$ and $f$ are
dimensionless and the dimension (length to $-3/2$) of $\phi_{23}$
compensate for the one length in the denominator of the last term.

The Fourier transform and the corresponding regularization then give
the momentum-space wave function, i.e.
\begin{equation}
\Psi _{\kappa _{0}}(\kappa ,\Omega _{\kappa }) = 
-\frac{2^{7/2}\pi ^{5/2}}{\kappa _{0}^{3/2}}
\frac{A}{\kappa ^{2}-\kappa _{0}^{2}}F(\Omega _{\kappa })  +
B \phi_{23}({\bf{p}}) \frac{4\pi }{q^{2}-q_{0}^{2}} f_{y}(\Omega _{q}) \;,
\end{equation}
where $q^{2} = \kappa ^{2} - 2 m B_{23}/\hbar^2$, $\phi_{23}({\bf{p}})$ is the
momentum-space wave function of the two-body bound state
$\phi_{23}$. These two channels correspond, respectively, to two close-lying
particles in a bound state far away from the third one, and three particles
all far away from each other. Thus, in the of large distances they do not 
interfere, and the resulting
momentum distribution is simply a weighted sum of the corresponding
distributions. The relative contributions of the two channels are
given by $|A|^{2}$ and $|B|^{2}$ correspondingly.

Generalization to describe decays into more than one two-body bound
state is formally straightforward, i.e. the corresponding
non-interfering asymptotic terms should simply be added.  This holds
for more than one bound state in the same two-body system as well as
for bound states in different two-body systems.

\subsubsection{One resonant two-body subsystem}

Instead of a bound state the decay via a two-body resonance is often
considered in interpretation and analysis of experiments.  Clearly the
narrower the resonance the more similarity to the case of two-body
bound states.  In any case the hyper-spherical expansion must
eventually converge, although the convergence can be too slow for a
reliable extraction of the asymptotic coefficients in eq.(\ref{asy})
from a numerical solution of the three-body problem.

However, in this case the (slowly convergent) two-body resonance
configuration can then be explicitly included into the asymptotics
while only the remaining (hopefully fast convergent) part is expanded,
i.e.
\begin{equation} 
\Psi _{\kappa _{0}}(\rho ,\Omega _{\rho })\stackrel{\rho \rightarrow \infty
}{\longrightarrow }A\frac{e^{+i\kappa _{0}\rho }}{\rho ^{5/2}}F(\Omega
_{\rho })+B\frac{e^{ip_{0}x}}{x}f_{x}(\Omega _{x})\frac{e^{iq_{0}y}}{y}
f_{y}(\Omega _{y}),
\end{equation}
where $\hbar ^{2}\kappa_{0}^{2}/(2m) = E_{0} -i \Gamma_{0} /2 $ is the
complex three-body energy, $\hbar ^{2}p_{0}^{2}/(2m)= E_{23}^{(0)} -i
\Gamma_{23} /2$ is the (complex) energy of the two-body resonance and
the remaining part is described by the complex momentum
$q_{0}^{2}=\kappa_{0}^{2} - p^{2}$. The precise definition of
$q_{0}^{2}$ arises from a constraint to be seen below.

The corresponding momentum-space wave-function is again given by the
regularized Fourier transform, i.e.
\begin{equation} \label{e95}
\Psi _{\kappa _{0}}(\kappa ,\Omega _{\kappa })=-A\frac{2^{7/2}\pi ^{5/2}}{
\kappa _{0}^{3/2}}\frac{F(\Omega _{\kappa })}{\kappa ^{2}-\kappa _{0}^{2}}+B
\frac{4\pi }{p^{2}-p_{0}^{2}}f_{x}(\Omega _{p})\frac{4\pi }{q^{2}-q_{0}^{2}}
f_{y}(\Omega _{q}).
\end{equation}
The momentum distribution is given by the absolute square of this
momentum-space wave-function. In the center of mass system we can
directly find the distribution of particle 3 arising from the
sequential emission via the two-body resonance. 

Absolute square of the last term in eq.(\ref{e95}) and use of the
energy conservation $\kappa^{2} = q^{2} + p^{2}$ (or $E = E_{1} +
E_{23}$) immediately gives the energy distribution for particle $1$
\begin{eqnarray}
 P(E_1) \propto  \int d E_{23}
 \frac{1}{[(E_{23}-E_{23}^{(0)})^2 + \Gamma_{23}^2/4]}
 \frac{1}{[(E_{23}+E_1 - E_{0})^2 + \Gamma_{0}^2/4]}  \nonumber \\ \propto
  \frac{1}{(E_1 - (E_{0}-E_{23}^{(0)})^2 + (\Gamma_{0}+\Gamma_{23})^2/4} \; ,
 \label{e100}
\end{eqnarray}
which states that the most probable energy of particle $1$ is $E_1 =
E_0-E_{23}^{(0)}$ and the width of the distribution is the sum of the
two and the three-body widths. Precisely this Breit-Wigner
distribution only arises when the $|q^{2}-q_{0}^{2}|^2$ in
eq.(\ref{e95}) is proportional to $(E-E_{0})^2 + \Gamma_{0}^2/2$,
i.e. given by the probability distribution in the initial three-body
state, which also is of Breit-Wigner form.  Thus the definition of
$q_{0}^{2}$ must involve $p^2$ and not $p_0^2$.

The same integration could of course be performed on the energy of
particle 1, but this would only give the two-body Breit-Wigner
distribution for $E_{23}$ whereas the measurements provide individual
energies, $E_2$ and $E_3$, for particles 2 and 3 in the center of mass
system.  To get the distributions of $E_2$ and $E_3$ involve trivial
but tedious kinematical transformations where also energies and
directions of particle 1 are required.

\subsubsection{Alternative real-coordinate procedure}

The relative energy distributions can be obtained from the angular
resonance wave function calculated without complex scaling.  We assume
that the system of particles is produced in an initial state for
example by a beta-decay process. We can then imagine the subsequent
decay as the time evolution of the initial non-stationary state. This
can be formulated as a time dependent coupled channels problem. It can
also be viewed intuitively as a particle described by time dependent
coordinates determined by classical equations of motion.  This should
be done with the appropriate initial amplitudes for all parts of the
initial wave function.  The hyperradius must vary from being very small
to infinitely large.  We increase $\rho$ until all particles are
outside the interaction ranges of all other particles. From then on
all distances scale as $\rho$ and all other coordinates remain
unchanged with time until $\rho = \infty$.

Energy conservation is maintained at large distances by converting the
potential energy into kinetic energy in the scaling degree of freedom,
i.e. by increasing the velocity $\dot{\rho}$ of the $\rho$-coordinate.
The wave function then evolves with all angular degrees of freedom
frozen eventually reaching the detectors placed far away.  The
absolute square of the angular wave function as function of
$\cos^2\alpha$ then provide the energy distributions simply because
the kinetic energy of particle 1 is given by the velocity of $y$,
i.e. $\dot{y} = \dot{\rho} \cos\alpha$.  Then the energy distribution
as function of the kinetic energy, proportional to $\dot{y}^2 \propto
\cos^2\alpha$, is the probability coordinate-space distribution as
function of $\cos^2\alpha$, apart from the phase space conversion from
$\alpha$ to energy, i.e. division by a factor proportional to
$dE/d\alpha \propto \sin(2\alpha)$.

This procedure is tempting since we only need to increase the maximum
value of $\rho$ in all the numerical implementations and plot the
wave function at that large distance.  For this to be accurate the
asymptotics has to be well described by the hyperspherical expansion
or the basis has to be very large and able to describe the necessary
large $\rho$-behavior.  However, when the basis functions have 
asymptotics different from one of the intermediate structures
the size of the basis needed to reach convergenge can easily become huge, 
making the procedure impractical.  This is not necessarily
easy to see in the numerical results where an increase of basis size
usually is rather expensive while the convergence could be extremely
slow.  The procedure is probably only directly useful for direct decay 
or for sequential decay via broad resonances. Otherwise the different
intermediate structure should be computed somehow and extrapolations
to large distances applied to each component individually.

An example to illustrate the present alternative formulation is
available in the schematic model discussed in details in
\cite{gar05a} where the widths but not the energy distribution were 
computed.  Assume that only the Coulomb potential is important and the
most probable path (ridge in the wave function) from small to large
distances can be described by scaling the hyperradius.  The
corresponding optimum path is defined by minimizing the WKB-tunneling
expression as function of different relative scaling parameters
$s_{ik} = r_{ik}/\rho$, where $r_{ik}$ is the distance between
particles $i$ and $k$. The path is given by $s_{ik}^{3} m_i Z_j =
s_{jk}^{3} m_j Z_i$ (see also \cite{kar04}), where $Z_i e$ is the
charge of particle $i$. We then arrive at the most probable value for
the energy division, i.e.
\begin{eqnarray} \label{e110}
E_k = E_{total} \bigg(1 + \big(\frac{m_k Z_k^2}{m_i Z_i^2}\big)^{1/3}
+ \big(\frac{m_k Z_k^2}{m_j Z_j^2}\big)^{1/3} \bigg)^{-1} \; ,
\end{eqnarray}
where $E_{total}$ is the total energy distributed among all the three
particles.  This expression is simple but not very accurate.  It also
only provides an estimate of the peak value. To compute the
distribution other paths must also be considered.  This is possible
but we shall leave this for a later discussion in connection with a
detailed treatment of the Coulomb interaction.

\section{Resonance wave functions}

The resonance wave function contains all information including that of
the relative energy distribution after the decay.  The calculations
must then first provide the corresponding three-body resonance states.
Second the large-distance behavior must be accurately extracted.  Due
to the different structures, these steps are not trivially completed
by use of only one method.  We briefly describe first the main
ingredients in our computations and the features of the wave function.
Second we explain how the asymptotic behavior is obtained in practice.

\subsection{Method}

We use the hyperspherical adiabatic expansion method combined with
complex scaling to obtain resonance wave functions. The coordinates are
defined in section 2 along with our basis functions in angular space,
i.e. the hyperspherical harmonics in each of the three Jacobi systems.
We solve the complex scaled Faddeev equations as function of
hyperradius \cite{nie01,fed03}.  The complex scaled coupled set of
radial equations are subsequently solved with the appropriate boundary
conditions, i.e. exponentially vanishing with increasing $\rho$ for
both bound states and resonances.  Thus here we assume that we do not
need to treat the Coulomb interaction explicitly at asymptotic large
distances. As shown in \cite{fed03}, the results obtained with this method 
agree well with some of the most common procedures, as for instance the 
complex energy method.

The energies are usually accurately determined in this method. The
same applies to the wave functions at small and intermediate distances
where the exponential fall-off still is not too restrictive. However,
we need the information at distances where the asymptotic limit is
reached, i.e. possibly at very large $\rho$ where the complex scaled
wave functions are very small.  Furthermore, more than one geometric
structure can be important at the same time, e.g. two different
spatially confined two-body configurations with the (different) third
particle far away.  This happens frequently with two identical
particles like neutrons and protons as constituent particles, because
the nucleon-core interaction must be sufficiently attractive to
produce a bound or resonating three-body system, and this implies that
such two-body configurations are favored.  To account simultaneously
for different two-body substructures, it is essential to use three
components as in the Faddeev decomposition adopted by us.  The same
efficiency can be achieved in a variational approach by allowing
Faddeev-like components in the trial wave function \cite{kam89}.  It is
much more difficult, if not impossible, to reach convergence with only
one component as in the hyperharmonic expansion method \cite{fed97}.

Even with three Faddeev components a large basis has to be
employed. To describe substructures inside one of the two-body
potentials (range $R_{eff}$) for large $\rho$ we need values of the
hyperpsherical quantum number $K$ up to a few times $\rho\sqrt{m} /
(R_{eff}\sqrt{\mu})$ where $\mu$ is the reduced mass of the two
particles. This is because $K/2$ is the number of nodes in the basis,
and details can only be described if a few nodes can be placed inside
the structure in question.  Thus, for nuclear systems, where $R_{eff}
\simeq 4$~fm, we need $K_{max}$ of at least $50$ to describe such
structures for $\rho \simeq 100$~fm.  Employing complex scaling
transforms resonances into states obeying the numerically easier bound
state boundary conditions.  The required basis size is larger,
essentially because the exponential fall-off moves to larger distances
with increasing scaling angle.  These estimates provide necessary
conditions for a reasonable description of two-body substructures
which in turn are necessary to describe the sequential decay mechanism.

\begin{figure}
\begin{center}
\vspace*{-1.1cm}
\epsfig{file=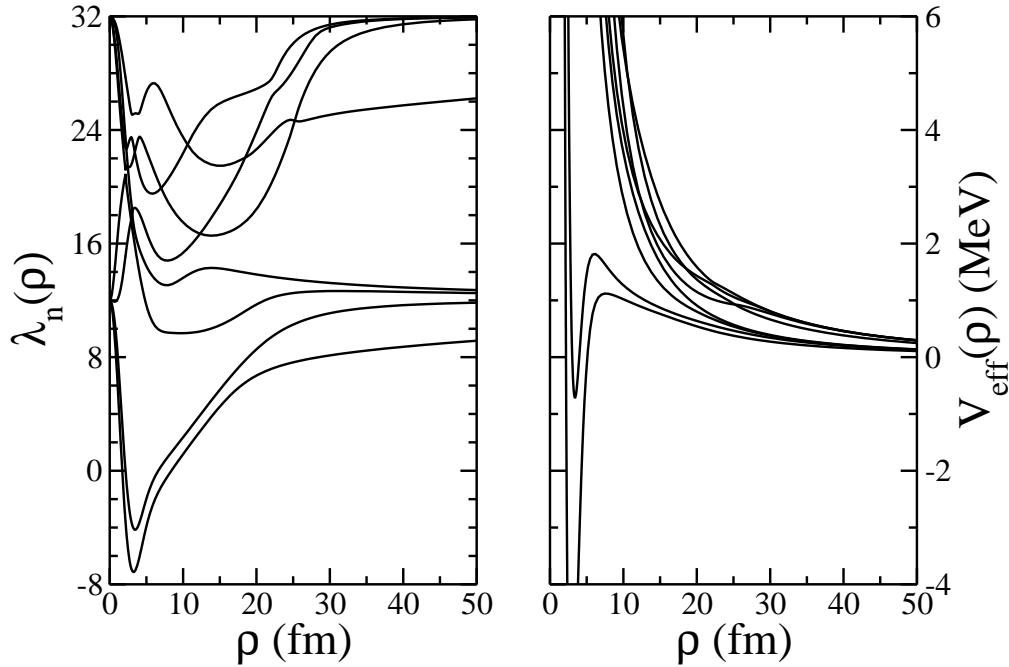,scale=0.5, angle=270}
\end{center}
\caption{ The real parts of the lowest 8 angular eigenvalues (left) and 
corresponding adiabatic potentials (right) as functions of $\rho$ for
the 2$^+$ states in $^{6}$He ($^{4}$He + n + n).  The scaling angle is
$\theta = 0.10$ rads. }
\label{fig1}
\end{figure}

The requirement of large $K$ to describe substructures must be
reconciled with the fact that an increase of $\rho$ towards infinity
results in convergence of the angular eigenvalues to the hyperharmonic
spectrum for free particles.  We show an example in Fig.~\ref{fig1}
where the imaginary parts are omitted as they both oscillate around
zero and approach zero at large $\rho$. The real parts of the angular
eigenvalues approach $K(K+4)$ as $\rho$ increases while the
corresponding potentials all approach zero faster than $\rho^{-2}$.
The attractive pockets in the eigenvalues at short distance disappear
in the potentials except for the two lowest where the negative values
remain as a prominent feature. The approach to the asymptotic values
is very fast except for the levels where $s$-waves contribute
significantly. The low energies favor these levels at large distances.

The related adiabatic wave functions approach the hyperspherical
harmonics.  The reason is that the regions in space, where the
short-range interactions are significant, are shrinking in size with
increasing $\rho$ relative to the total space available. The radial
extension of these regions, responsible for two-body correlations,
decrease with $\rho$ as $1/\rho$.  The interactions are non-vanishing
in smaller and smaller regions.  Consequently they become less and
less important for both energies and wave functions. Thus, the basis
size has to increase with $\rho$ in order to allow a description of
the two-body substructures or equivalently of sequential decay, but
the lowest adiabatic potentials approach the free solutions.  The
basis size in practice always has to remain finite and the
substructures eventually become impossible to describe in this way.

The interactions have to be chosen to reproduce the pairwise
low-energy scattering properties of the three particles appearing in
the final state.  Clearly we must accurately include all the partial
waves necessary to describe the quantum numbers of the decaying
resonance. However, even with a sufficiently large basis the
three-body system does not necessarily appear with the correct energy
and width.  In fact, there may not even be an attractive region at
small distances as required to produce a resonance of finite width.
This could occur when we are dealing with a many-body resonance
without traces of any three-body cluster structure.  Nevertheless, a
meaningful computation can be carried out of the energy distributions
emerging after a three-body decay.

The philosophy is the same as for $\alpha$-emission where the inner
part of the effective potential is replaced by an attractive square
well with a depth adjusted to reproduce the resonance energy.  The
resulting barrier is then used to derive the width, usually in the WKB
approximation. We generalize this concept to the adiabatic potentials,
i.e. we add a three-body potential of short range in the
hyperradius. It is intended to describe interactions beyond the
two-body level such that the three-body system has a resonance at the
desired energy.  By doing this we have substituted the possibly
complicated many-body structure at small distance with the three-body
cluster structure resulting from an effective potential, which in turn
also provides the correct boundary conditions for a three-body
decaying resonance.  This principle was introduced for fine-tuning in
the first calculation with the correct boundary conditions of the
three-$\alpha$ decay of the second $0^+$-state in $^{12}$C
\cite{fed96}. It has later become the standard procedure to adjust
three-body energies without significant changes of the underlying
substructure \cite{nie01}.

\subsection{Important features}

The radial solution is often strongly dominated by one or two of the
lowest adiabatic components at small distance where the relative
probability is large.  This is because all three short-range two-body
interactions contribute simultaneously and the result is the
energetically most favored three-body resonance structure consistent
with the boundary conditions.  As $\rho$ increases, at least one
particle has to move away from the other leaving at most one
non-vanishing two-body interaction. Coherent contributions from
different of these configurations are possible and sometimes even
favored.  At large distance, where the energy distribution is
determined, several more adiabatic potentials are often needed.  The
couplings due to the Coulomb interaction would normally increase the
necessary number of potentials.

\begin{figure}
\begin{center}
\vspace*{-1.1cm}
\epsfig{file=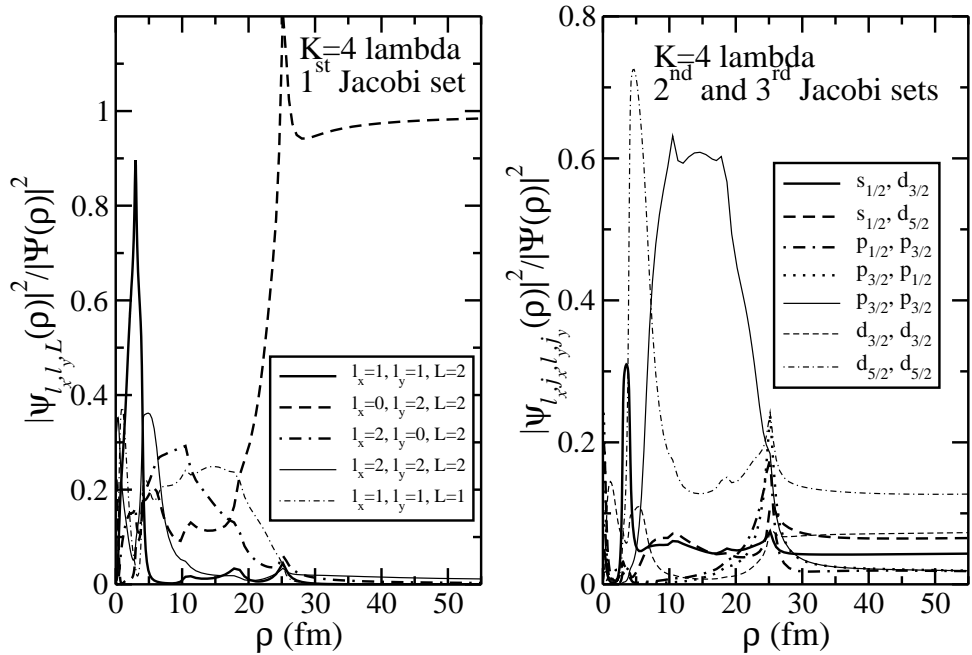,scale=0.5, angle=270}
\end{center}
\caption{The fraction of different components in the fifth adiabatic 
potential for $\theta=0.10$ rads as function of $\rho$ for $^{6}$He(2$^+$).
The angular eigenvalue corresponds to $K=4$ at large $\rho$, see
Fig.~\ref{fig1}.  The angular momenta are specified by $\ell_x$,
$j_x$, $\ell_y$, $j_y$, and $L$.  Left: $x$ refers to the two-neutron
system and $y$ to its center of mass motion relative to the
$\alpha$-particle.  Right: $x$ refers to the neutron-$\alpha$ system
and $y$ to its center of mass motion relative to the other neutron.
We give the $(x,y)$ components on the figure as $\ell_{j}$.  }
\label{fig2}
\end{figure}

It is established \cite{nie01} that the adiabatic potentials of lowest
energy at large $\rho$ are related to configurations with relative
$s$-waves between the two closest particles.  This is the basis for
the Efimov effect \cite{efi70}.  If these large-distance
configurations differ from the resonance structure at small $\rho$,
possibly with higher partial waves, the lowest angular wave function
must change its structure accordingly as $\rho$ increases.  In
\cite{gar05b} we showed the structure for $^{6}$He(2$^+$) for the
lowest eigenvalue which approach the $K=2$ value at large $\rho$. In
Fig.~\ref{fig2} we show the results for the similar eigenvalue
approaching the $K=4$ level for large $\rho$.  The pronounced and
rapidly changing structure is qualitatively similar to the lower-lying
$K=2$ level, i.e. dominated by $p_{3/2}-p_{3/2}$ neutron-core
structure at small $\rho$ and by $s$-waves between the two neutrons at
large $\rho$.  Essentially all other allowed components contribute
with equally small amounts. 

These rather dramatic changes imply that it is crucial to include all
adiabatic potentials with significant couplings to those dominating
the structure at small hyperradii.  This is simply because the
couplings are responsible for changing the radial weights of the
different adiabatic components as function of $\rho$, e.g. no
couplings imply the same occupation independent of $\rho$.  On the
other hand, each of the angular wave functions related to the adiabatic
potentials are themselves functions of $\rho$, sometimes rapidly
varying as seen in Fig.~\ref{fig2}.  In principle the non-diagonal
couplings could be vanishingly small and all change of structure would
be described by the lowest adiabatic wave function.  However, this is
rather unlikely because the couplings are defined as matrix elements
of first and second radial derivatives of the angular
wave functions. Thus, radial couplings between rapidly changing angular
structures are inevitable.  

\begin{figure}
\begin{center}
\vspace*{-1.1cm}
\epsfig{file=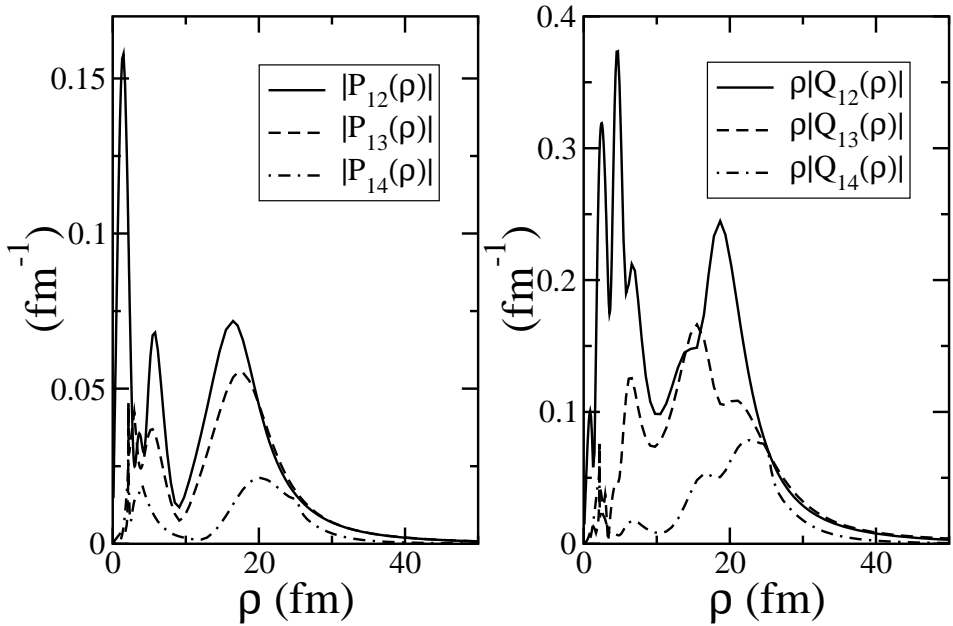,scale=0.5, angle=270}
\end{center}
\caption{The coupling potentials between the four dominating adiabatic 
levels for $\theta=0.10$ rads shown as functions of $\rho$ for 
$^{6}$He(2$^+$).  The
first and the fourth levels have similar quantum numbers but approach
the $K=2$ and $4$ levels, respectively.  To show the first ($P$) and
second ($Q$) order coupling potentials in the same units (fm$^{-1}$)
we multiply $Q$ by $\rho$. (The energy unit is restored in the
coupling potentials by including the omitted factor, i.e. $\hbar^2
Q/(2m)$, $\hbar^2 P/(2m)\partial/\partial \rho$). }
\label{fig3}
\end{figure}

In Fig.~\ref{fig3} we show how the strongly varying coupling
potentials can be related to the changing angular structure seen in
Fig.~\ref{fig2}.  The peaks are most pronounced when a crossing is
avoided and the two levels switch characteristics \cite{nie01}.  The
rather confusing coupling picture is crucial for the asymptotic
behavior of the wave function at large distance where the energy
distribution is determined. The second order terms are substantially
larger than the first order couplings but all vanish at large $\rho$.
Thus the numerical computations in the present case must extend at
least beyond $\rho \approx 40$~fm where the couplings have reached
very small values.  This is about two times the largest scattering
length which defines the distance of convergence towards asymptotic
values.

We have now established two important but competing effects, which
determine the three-body resonance structure from small to large
values of $\rho$.  In the extreme, the structure can either remain
unchanged by climbing correspondingly up on the adiabatic potentials,
or the structure can change to follow that of the lowest-lying
adiabatic wave function.  A compromise between following the
energetically most favored configuration and the resistance to a
change of structure therefore must be reached.  The combination of
these effects determine the relative population of the different
components in the radial solution, which in turn determines the
observable energy distribution.  The couplings are more important here
than for widths, energies and small-distance wave functions.  They
must be accurately computed to provide the energy distribution.

The structure of the resonance wave function at large $\rho$ could
remain unchanged and only exhibit a simple scaling behavior
proportional to $\rho$. This is typical of direct decay. The
wave function could also have large probability for finding two
close-lying particles where the hyperradius mainly changes by moving
the third particle as $\rho$ increases.  This is typical of sequential
decay via more or less stable two-body configuration, e.g. sequential
decay through two-body resonances.  The intermediate configuration
does not necessarily need a confining barrier, but could be provided
by low-lying two-body virtual $s$-states \cite{gar05c}.  Mixtures of
all types can occur giving rise to the description of decay properties
as fractions proceeding via individual two-body configurations.  All
these structures can be computed by use of our method, although
convergence for the Coulomb interaction is more difficult.

The best strategy to get reliable results is not obvious, because the
brute force method of increasing basis size and hyperradius until
convergence is reached may be beyond any reasonable computer effort.
The indecision is related to the requirement of an increasing basis
with increasing $\rho$, which means that a smaller $\rho$ and a
smaller basis could provide a better description with much less
effort. In other words a convergence may be reached in a region of
$\rho$-values for a moderate basis size.  This convergence would be
destroyed as $\rho$ is allowed to increase because the basis size
cannot follow. The convergence can possibly be reached faster by
extrapolation of the observable distribution by use of a known or
anticipated dependence of $\rho$ and basis size \cite{res99}.
Different parts of the wave function may extrapolate differently.  The
most efficient choice depends on the (mixtures of) decay mechanisms
which therefore has to be determined first.  Therefore the first step
is to compute the structure of the resonance wave functions as
discussed in \cite{gar05b}.

\section{Realistic numerical illustration: $^{6}$He($2^+$)}

Nuclear three-body decay without complications of the Coulomb
interaction must involve emission of two neutrons.  The decaying
states do not have to be three-body structures although such
two-neutron halos are available and rather well studied. The most
obvious case is the established $2^+$ resonance in $^{6}$He which is
formed by the same neutron-core components as in the ground state. 

Without Coulomb interactions the computations should quickly lead to
the desired energy distributions.  However, even short-range
interactions can present difficulties as highlighted by the intricate
description needed for the Efimov effect \cite{nie01}.  Both the
$\alpha$-neutron and the neutron-neutron interactions are previously
employed in ground state computations \cite{nie01}.  

We follow the
procedure outlined in the preceding sections. Different prescriptions 
are possible to implement the Pauli principle \cite{gar99}, all of
them providing indistiguishable angular wave functions at large distances.
We adjust the three-body potential to give the correct resonance energy.  
This only requires marginal fine-tuning. In total 1132 hyper-spherical 
harmonics are used in the expansion (\ref{eq21}), and the maximum value of
$K$ is 200 for the most relevant partial wave components and never smaller 
than 40.  The resonance wave function is then
available as function of the hyperspherical coordinates. A complex scaling
angle of 0.10 rads is enough to produce an exponentially vanishing with 
increasing $\rho$ wave function for the resonance. A different scaling 
angle, where the numerical calculations have converged, produce the same 
results.

\subsection{Resonance structure}

We already showed the angular eigenvalues and the adiabatic potentials
in Fig.~\ref{fig1}.  The probability distribution arising from only
the lowest potential was shown in \cite{gar05b}. The structure changes
from peaks at small $\rho$ corresponding to $\alpha$-neutron
$p_{3/2}$-structure to a probability with one broad peak corresponding
to comparable distances between all three particles. This reflects the
change of structure of this angular wave function from small to large
$\rho$ as seen in details in Fig.~\ref{fig2}. Eventually the lowest
hyperharmonic function with $K=2$ is approached.  This indicates in
itself a direct decay mechanism. However, in this case the lowest
potential provides rather misleading results.

\begin{figure}
\begin{center}
\vspace*{-1.1cm}
\epsfig{file=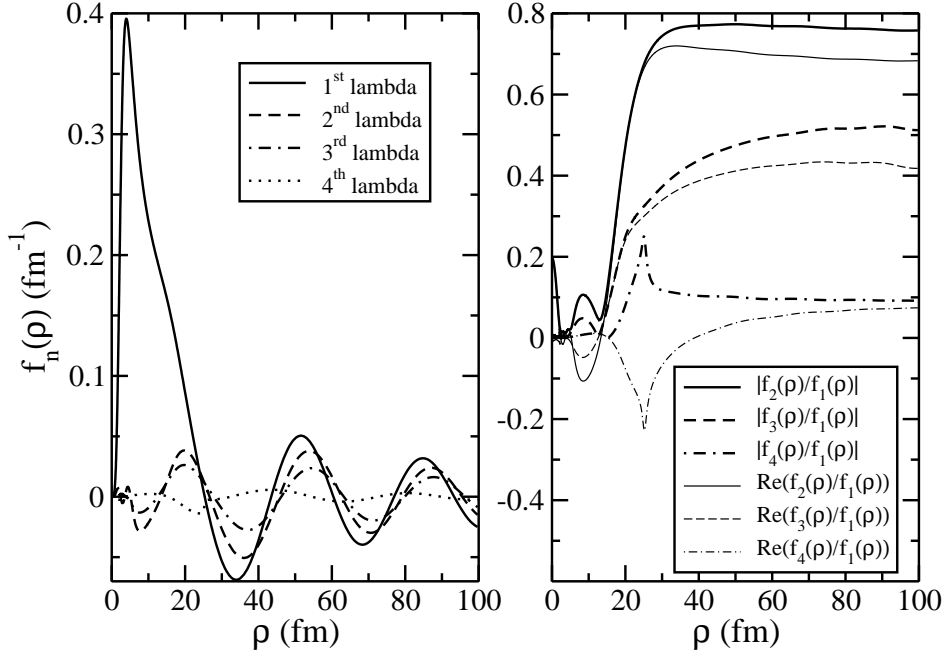,scale=0.5, angle=270}
\end{center}
\caption{The radial wave functions (left) and the absolulte values and
real parts of their relative sizes (right) corresponding to the four 
dominating adiabatic potentials for $\theta=0.10$ rads as functions 
$\rho$ for the $^{6}$He(2$^+$) resonance.}
\label{fig4}
\end{figure}

The rapidly changing structure seen in Fig.~\ref{fig2} at around $\rho
\approx 20$~fm could easily lead to occupation of higher-lying
levels. These occupation probabilities are functions of $\rho$ and
simply found as squares of the radial wave function obtained by solving
the coupled set of radial equations.  The results are shown in
Fig.~\ref{fig4} for the lowest adiabatic components. At small $\rho$
the lowest potential is totally dominating but as $\rho$ increases the
lowest three components contribute with comparable amplitudes.  All
radial wave functions vanish by oscillating around zero with decreasing
amplitudes.  The relative sizes are more clearly seen in the right
hand side of Fig.~\ref{fig4}. After the transition around 20~fm the
individually very small radial amplitudes stabilize on relatively
constant finite ratios.  The square of these give the relative
weights, i.e. reduced compared to the first component by about 0.6,
0.25, 0.01 for the second, third and fourth potential, respectively.
The transition to stable ratios is consistent with the disappearance
of the coupling terms shown in Fig.~\ref{fig3}.

\begin{figure}
\begin{center}
\vspace*{0.1cm}
\epsfig{file=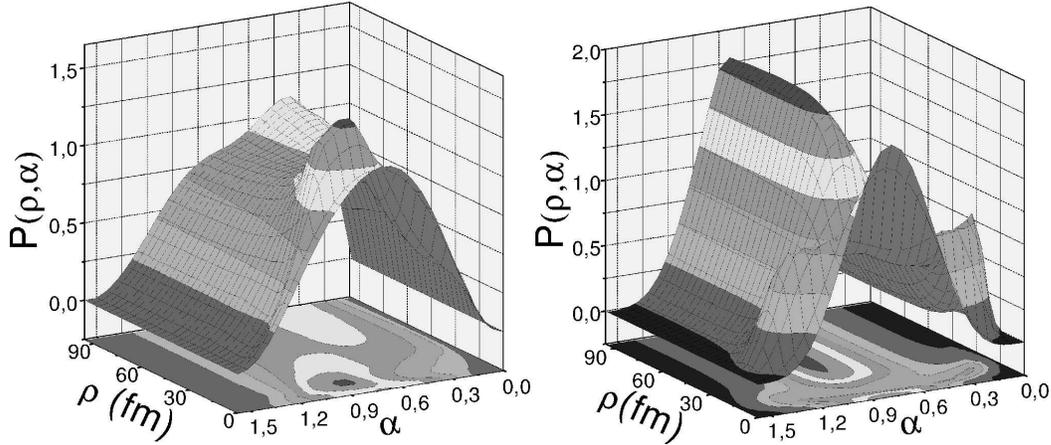,scale=0.8,angle=0}
\end{center}
\caption{The probability distribution for $^{6}$He(2$^+$) including
the lowest 8 adiabatic potentials as function of hyperradius $\rho$
and hyperangle $\alpha$ related to the distance by $r_{ik} \propto
\rho \sin \alpha$, i.e. the distance between either the one neutron and 
core $r_{nc}$ (left) or the two neutrons $r_{nn}$ (right).  }
\label{fig5}
\end{figure}

The total probability distribution in Fig.~\ref{fig5} are quite
different from that of the lowest eigenvalue.  At large $\rho$ the
probability now peaks at a smaller distance between the two neutrons
and correspondingly the $\alpha$-neutron distance is increased.  Still
fairly broad distributions remain.  The decay mechanism indicated by
this structure is now instead of direct rather a mixture between the
preferred sequential decay via a neutron-neutron intermediate
configuration and a smaller direct component.

\subsection{Energy distributions}

\begin{figure}
\begin{center}
\vspace*{-1.1cm}
\epsfig{file=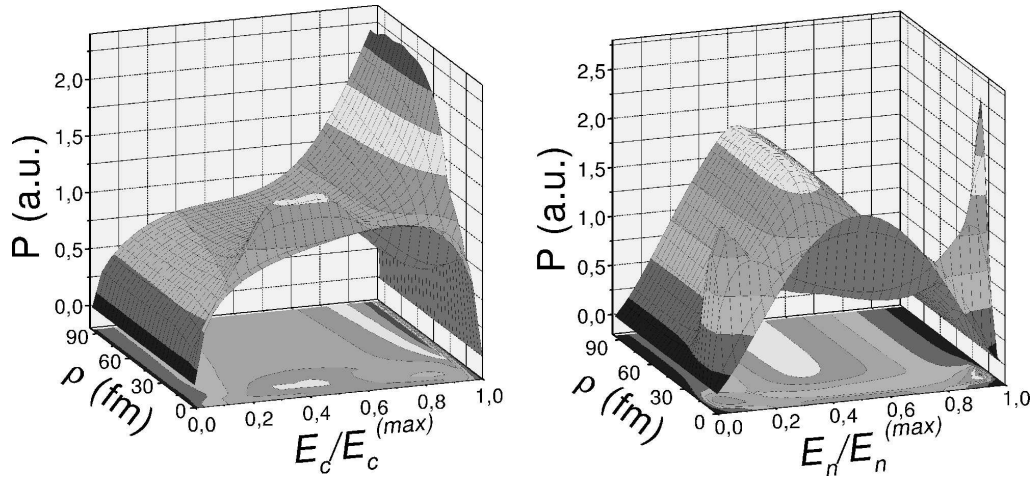,scale=0.8,angle=0}
\end{center}
\caption{The energy distributions of neutrons (right) and 
$\alpha$-particles (left) after decay of $^{6}$He(2$^+$) for $\theta=0.10$
rads. The three-dimensional plot show the dependence on $\rho$ with 
inclusion of 8 adiabatic wave functions. The maximum energies are
$(m_{\alpha}+m_n)/(m_{\alpha}+2m_n) E_0$ and $2m_n /(m_{\alpha}+2m_n)
E_0$ for the neutron and the $\alpha$-particle, respectively. Here
$E_0$ is the energy of the decaying resonance. }
\label{fig6}
\end{figure}

Reliable computation of the energy distribution requires numerically
converged results in an appropriate region of $\rho$-values.  The
energy distributions are shown in Fig.~\ref{fig6} as functions of
$\rho$ for a sufficiently large number of adiabatic potentials.  The
resemblance with the probability distribution is not surprising since
only the volume element has been changed. The observable distribution
is the cut for constant, and sufficiently large, $\rho$ where
convergence has been reached as function of basis size.  The neutron
energy distribution has two peaks for small $\rho$ corresponding to
the geometric configurations of one neutron close to the
$\alpha$-particle and the other neutron further away. This is
reflected in the peak in the $\alpha$-spectrum at intermediate
energies corresponding to the same geometric configurations.

The structure changes with $\rho$ into a broad peak at intermediate
energies for the neutron spectrum, and one peak very close to the
maximum energy for the $\alpha$-spectrum.  This is the fingerprint of
sequential decay via emission of the $\alpha$-particle followed by
decay of an intermediate two-neutron structure.  This is easily
visualized as the two-body decay process where the $\alpha$-particle
receives maximum energy when the two neutrons move together in the
opposite direction.  In the subsequent decay each neutron then must
share the remaining energy which leads to an intermediate energy
between zero and the maximum value.

This inferred decay mechanism is perhaps counter-intuitive because
stable intermediate configuration of two neutrons do not exist neither
as bound states nor as resonances. It would be much more acceptable
with the $\alpha$-neutron $p_{3/2}$-resonance as the intermediate
structure.  However, one characteristic feature of the neutron-neutron
interaction is the low-lying virtual $s$-state which simply means that
there is a substantial $s$-wave attraction. Apparently this is
decisive for the decay process where the two neutrons end up by moving
essentially in the same direction, and then necessarily guided by the
attraction. The interesting point is maybe that this is not the way
they started out at small distance in the spatially confined part of
the wave function. This change of structure with hyperradius is in a
sense reflecting the dynamic character of the decay process.  At
distances larger than the scattering length the short-range
interactions are negligibly small, the wave function changes are
completed and the asymptotics are established.

\begin{figure}
\begin{center}
\vspace*{0.0cm}
\centerline{\psfig{figure=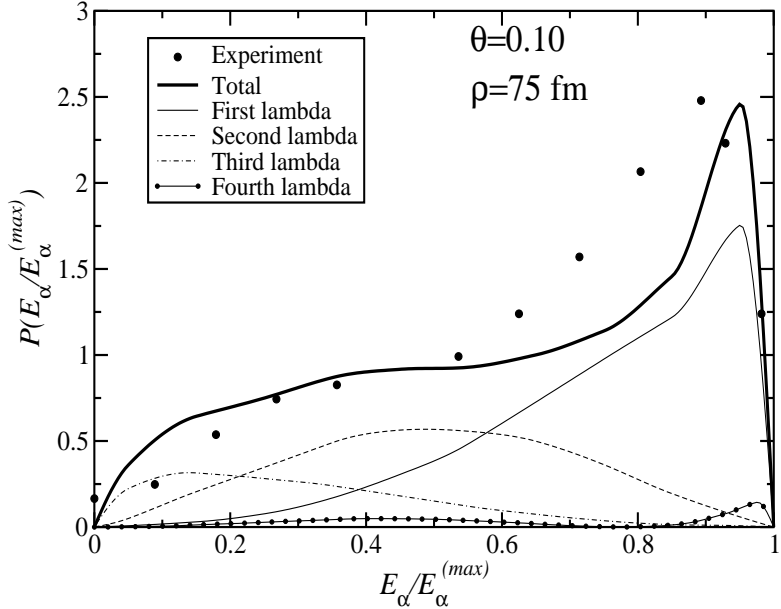,height=10.0cm,width=8.0cm,%
bbllx=2.8cm,bblly=0.6cm,bburx=20.4cm,bbury=24.5cm,angle=270}}
\end{center}
\caption{ The energy distribution of the $\alpha$-particle after decay 
of the $2^+$-resonance in $^{6}$He. The scaling angle is $\theta
=0.10$ rads and $\rho=75$~fm where convergence is reached.  The points are
extracted from the measurements in
\cite{dan87}. Contributions from the 4 dominating adiabatic potentials are
shown individually.  }
\label{fig7}
\end{figure}

The microscopic structure of the energy distributions can be studied
by dividing into contributions from the individual adiabatic
potentials as seen in Fig.~\ref{fig7} for the emitted
$\alpha$-particle.  The total distribution remains essentially
unchanged if more than the four dominating potentials are included.
Each contribution has its own characteristic feature.  The first has a
peak close to the maximum energy, i.e. resembling $\alpha$-emission
from a neutron-neutron $0^+$-state. The second has a peak at
intermediate energy, i.e. resembling sequential decay by the
$\alpha$-neutron resonance.  The third has a peak at small energy,
i.e. resembling $\alpha$-emission from an excited neutron-neutron
$2^+$-state. In addition the fourth potential also gives a small
contribution with maxima at intermediate and maximum energy. The size
of about 1\% cannot be seen in the total distribution. However, this
eigenvalue has the same angular momentum quantum numbers as the first
level. Therefore the non-diagonal interference term would be about
10\% of the total contribution.  It turns out that the interference
essentially is destructive and responsible for the almost flat region
at intermediate energies.

The decay mechanism is then not simple although understandable in
terms of our formulation.  The main contribution is decay via the
virtual $s$-state and the second is from direct decay.  The third
mechanism is produced by the coupling to the higher-lying state taking
place at relatively small $\rho$.  This populates the level eventually
approaching the $K=4$ hyperspherical level at large $\rho$.  The
interference with the dominating contribution then leads to the total
distribution.  The division into these different contributions of
direct and sequential is to some extent artificial but perhaps useful
in connection with the experimental analysis.

\begin{figure}
\begin{center}
\vspace*{0.0cm}
\centerline{\psfig{figure=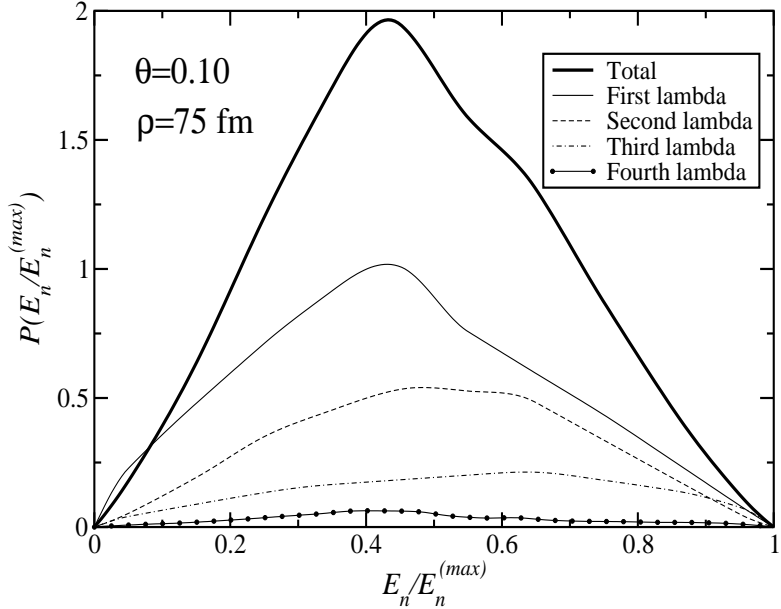,height=10.0cm,width=8.0cm,%
bbllx=2.8cm,bblly=0.6cm,bburx=20.4cm,bbury=24.5cm,angle=270}}
\end{center}
\caption{ The same as  Fig.~\ref{fig7} for the neutrons emerging after 
decay of the $2^+$-resonance in $^{6}$He.  }
\label{fig8}
\end{figure}

The mixture of all these contributions leads to the total distribution
which has the right features but without precise reproduction of the
high-energy peak, see Fig.~\ref{fig7}.  Experimental resolution would
not improve very much because either the peak gets broader and lower,
or it gets higher and narrower.  The discrepancies can originate from
the presence of the target and the reaction mechanism itself as well as
from contributions to the experimental points from other than resonance
decays.  The experiment selects the window of energies around the
$2^+$ resonance position in the reaction $^7$Li($^2$H,$^3$He)$^6$He$^*$.
This necessarily includes some background which perhaps has a
different energy distribution than the $2^+$ resonance we investigated
in the present calculations.  We find a distribution where the
two-neutron virtual $s$-state dominates whereas the measurements are
broader as expected from non-resonance decays. In this work we focus on 
the decay of "populated" resonances, and an appropriate description of this 
reaction goes beyond the scope of the paper.

An attempt to understand the distribution was published soon after the
experiment in \cite{dan87}. The measured distribution was fitted by a
linear combination of the lowest hyperharmonic functions of $K=2$ and
4. The conclusion was that a substantial $K=4$ component is needed to
reproduce the experiment. The decay mechanism dominated by the
neutron-neutron virtual $s$-state was abandoned in favor of the $K=4$
component. This phenomenological analysis provides a good fit even for
energies above the maximum allowed by the resonance energy.  The
decaying resonance wave function does not enter anywhere.  The
significance is not easy to interpret in terms of decay mechanisms as
attempted in the present work.

The neutron energy distribution is not measured but for future
comparison we show our prediction in Fig.~\ref{fig8}.  The division
into different adiabatic components show that the broad total
distribution centered around an intermediate energy is obtained by
adding several qualitatively similar contributions.  The different
mechanisms would all produce most likely energies around half the
maximum value.  To distinguish it is therefore necessary to measure
both $\alpha$-particles and neutrons after the decay.

\subsection{Dependence on scaling angle}

It is instructive to investigate the dependence of the distributions
on the choices of $\rho$, $\theta$ and basis size. The
$\rho$-dependence is already indicated in Fig.~\ref{fig6} where the
distributions are very stable as soon as $\rho$ is larger than about
$50$~fm.  However, this stability does require a sufficiently large
basis which at least up to $100$~fm still can be handled in
modest-size computers.  It is also clear that a finite basis cannot
accurately describe the solutions when $\rho$ increases towards
$\infty$. Then the angular solutions approach the hyperharmonics but a
large basis is still required to reproduce the structures at small
distances between pairs of particles.  Eventually this becomes
impossible.  The many basis functions cancel each other at larger
distances.

At intermediate distances, where the basis is sufficiently large, the
resonance wave functions are independent of $\rho$.  For the radial
solution this is seen in the right part of Fig.~\ref{fig4}, where for $\rho$ 
larger than about 50~fm, the ratio between the different radial components
is approximately constant. The energy distributions are mainly dominated 
by the absolute squares of these ratios, although when different adiabatic
components interfere also the real parts of these complex ratios may 
contribute individually. This behavior of the radial ratios is responsible
for the stable behaviour of the energy distributions for sufficiently large 
values of rho. The constant behavior of the radial ratios also implies that
the radial wave functions have already reached asymptotics as given in 
eq.(\ref{eq22}) for all the channels, and therefore the distributions are
independent of the scaling angle. However, the latter conclusion is based on 
an assumption of analyticity of the angular solutions as function of
$\theta$.  When this scaling angle is changed corresponding to a
rotation across a singularity like a two- or three-body resonance the
continuity is broken and the solutions change as well.

This is especially clear when we compare two solutions with $\theta$
smaller and larger than the angle corresponding to a two-body
resonance.  For the large $\theta$ one angular eigenvalue changes
character and increases towards infinity as $\rho^2$, see
\cite{fed03}.  This qualitative change of behavior necessarily causes
a change of the angular wave functions because the upgoing eigenvalue
at large distances fully describes the properties of the two-body
resonance.  These features were distributed over several wave functions
for the small $\theta$-value.

In between singularities the individual angular solutions are
independent of both $\rho$ and $\theta$.  This may not be an apparent
feature of the numerical solutions because the basis has to be
sufficiently large for a complete description.  As $\theta$ increases
the effective ranges of the two-body interactions also increase and
the stable region is pushed to larger $\rho$-values.  This means that
the minimum basis size has to increase with $\theta$.

\section{Summary and conclusions}

We formulate a method to compute the energy distribution of three
particles emerging after three-body decay of a many-body resonance.
The complex energy of a resonance corresponds to a pole in the
momentum-space wave function which has an absolute square of the form
as Breit-Wigner shape multiplied by a smoothly varying function.  In
coordinate-space this form corresponds to a large-distance asymptotic
wave function consisting of only outgoing waves. We show formally by
Fourier transformation that the coordinate-space asymptotic angular
dependence determines the energy distribution by substituting momentum
directions for the conjugate coordinate directions. For this the
divergent Fourier integral is regularized by the Zeldovich
prescription.

For two-body decay the energy distribution is trivially given by the
Breit-Wigner distribution of the initial resonance.  Energy
conservation is taking care of everything else.  For three-body decay
the total energy can be distributed continuously among the three
particles.  We show that the resonance decay results in distributions
obtained from the large-distance angular behavior of the coordinate
wave function.  The asymptotic behavior can correspond to either
genuine three-body structures or two-body substructures for example
corresponding to two-body resonances or configurations favored by
substantial attraction as for virtual states. Also virtual population
of two-body intermediate substructures is allowed as an appropriate
asymptotic behavior with a resulting special energy distribution.  The
different asymptotics characterize the different decay modes used in
analyses of experimental data.   Different modes can co-exist.

We illustrate by the decay of the $2^+$-state in $^{6}$He. The
practical computations employ the hyperspherical adiabatic expansion
combined with the complex scaling method. We discuss how a large
hyperradius necessarily must be accompanied by a large basis.
Convergent results may then be obtained with less efforts at moderate
hyperradii and moderate basis sizes. For convergence it is crucial to
have all three Faddeev components, and especially if all decay
mechanisms simultaneously should be included in the theoretical
formulation. The wave function undergoes dramatic changes from small
distances, where the resonance properties usually are determined, and
large distances from which the energy distributions emerge. The
reasons for these structural changes are that the small distance
behavior is determined by the two-body resonance substructures,
whereas the large-distance behavior is determined as a competition
between two effects, i.e. the energetically favored configuration of
smallest two-body angular momentum with attractive two-body potentials
and maintaining the same structure as at small distances in
higher-lying levels.

In conclusion, theoretical interpretation of the simplest nuclear
three-body decay without Coulomb interactions is already rather
complicated.  It is then advisable to test any given method on these
systems.  The accuracy of computations of the more complicated
decaying charged systems can then be judged. This is important since
almost all nuclear three-body resonance decays involve charged
systems.  The goal is to interpret the soon-to-come accurate
experimental correlation data for three-body decays of charged
systems.

\end{document}